# Magnetic properties of the double perovskites LaPbMSbO₆ (M = Mn, Co and Ni)

D. G. Franco[1,2], R. E. Carbonio[1] and G. Nieva[2,3]

[1]INFIQC-CONICET, Dpto. de Físico-Química, Facultad de Ciencias Químicas, Universidad Nacional de Córdoba, Ciudad Universitaria. X5000HUA Córdoba, Argentina
[2]Laboratorio de Bajas Temperaturas. Centro Atómico Bariloche – CNEA. 8400 Bariloche, Río Negro, Argentina.
[3]Instituto Balseiro, CNEA and Universidad Nacional de Cuyo. 8400 Bariloche, Río Negro, Argentina.

New double perovskites LaPbMSbO₆, where $M^{2+}$ = $Mn^{2+}$, $Co^{2+}$, and $Ni^{2+}$, were synthesized as polycrystals by an aqueous synthetic route at temperatures below 1000 ºC. All samples are monoclinic, space group P2₁/n, as obtained from Rietveld analysis of X-ray powder diffraction patterns. The distribution of $M^{2+}$ and $Sb^{5+}$ among the two octahedral sites have 3% of disorder for $M^{2+}$ = $Ni^{2+}$, whereas for $M^{2+}$ = $Mn^{2+}$ and $Co^{2+}$ less disorder is found. The three samples have an antiferromagnetic transition, due to the antiferromagnetic coupling between $M^{2+}$ through super-superexchange paths $M^{2+}$-$O^{2-}$-$Sb^{5+}$-$O^{2-}$-$M^{2+}$. Transition temperatures are low: 8, 10 and 17 K for $Mn^{2+}$, $Co^{2+}$, and $Ni^{2+}$ respectively, as a consequence of the relatively long distances between the magnetic ions $M^{2+}$. Besides, for LaPbMnSbO₆ a small transition at 45 K was found, with ferrimagnetic characteristics, possibly as a consequence of a small disorder between $Mn^{2+}$ and $Sb^{5+}$. This disorder would give additional and shorter interaction paths: superexchange $Mn^{2+}$-$O^{2-}$-$Mn^{2+}$.

## I. INTRODUCTION

DOUBLE perovskites of general formula A₂BB′O₆ and related materials are among inorganic compounds of great interest because of the large variety of physical properties that they show, based on the different elements A, B, and B′ that can be accommodated in the structure. Although some materials with double perovskite structure have been known for long [1], there was a revival of interest in this kind of systems due to the discovery of half metallicity and colossal magnetoresistance in Sr₂FeMoO₆ [2], with applications in spintronics [3].

In this work we report the synthesis, crystal structure and magnetic properties of the new double perovskite family LaPbMSbO₆, where $M^{2+}$ = $Mn^{2+}$, $Co^{2+}$, and $Ni^{2+}$.

## II. EXPERIMENTS

Because Pb is a highly volatile element, polycrystals of the new double perovskites LaPbMSbO₆ with $M^{2+}$ = $Mn^{2+}$, $Co^{2+}$, and $Ni^{2+}$ were synthesized at low temperatures by an aqueous synthesis route, which allows a high mixing of the reactants and lowers the reaction temperature with respect to the solid state method. For $M^{2+}$ = $Co^{2+}$ and $Ni^{2+}$, a 10% excess of lead was necessary to reach a high purity phase and the final annealing was performed at 900 ºC in air. For $M^{2+}$ = $Mn^{2+}$, 800 ºC was enough to obtain a single phase sample, but an inert atmosphere was necessary to stabilize $Mn^{2+}$. The detailed synthesis [4] will be published elsewhere.

The X-ray powder diffraction patterns (XRPD) were measured in a PANalytical X´Pert PRO diffractometer, with Bragg-Brentano geometry and Cu K$_\alpha$ radiation, λ = 1.5418 Å. The crystal structure was refined by the Rietveld method [5] using FULLPROF program [6].

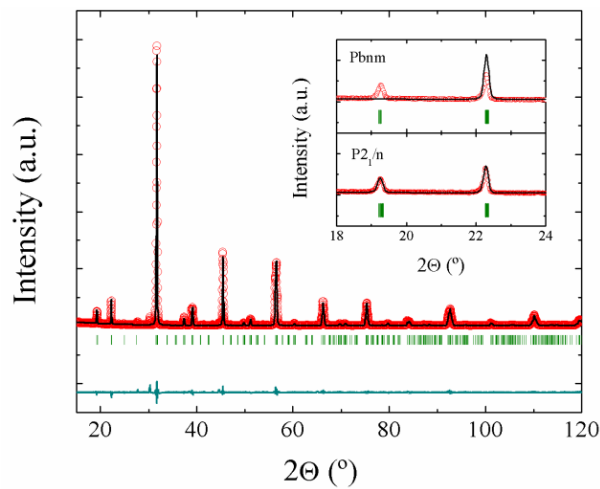

Fig. 1. Rietveld refinement of the XRPD data for LaPbNiSbO₆ in the space group P2₁/n. $\chi^2$ = 2.97, R$_{wp}$ = 13.1, R$_B$ = 4.00. Inset: detail of the low angle refinements for simple Pbnm, and double P2₁/n perovskite models, validating the superstructure. Experimental (circles), calculated (line), difference (bottom line) and Bragg reflections (vertical bars).

**FIG. 1 HERE**

Magnetic measurements were obtained in a commercial SQUID magnetometer between 5 and 380 K and -5 to 5 T. The magnetization as a function of temperature ($M$ vs $T$) was measured using a ZFC-FC procedure (i.e. cooling with zero applied field or with a finite applied field).

## III. RESULTS AND DISCUSSION

The perovskites LaPbMSbO$_6$ crystallize in the monoclinic space group P2$_1$/n, according to the Rietveld refinements, as shown for LaPbNiSbO$_6$ in Fig. 1. The unit cell is pseudo orthorhombic ($\beta \sim 90°$) for the three samples and whereas for the cuboctahedral A ions (La$^{3+}$ and Pb$^{2+}$) there is only one site, for the octahedral B ions (M$^{2+}$ and Sb$^{5+}$) there are two crystallographic sites. Therefore the perovskites are double. These two sites allow the order of M$^{2+}$ and Sb$^{5+}$ ions in a rock salt pattern. A double perovskite structure is the best choice because the observed superstructure peaks at low angle are not justified by an orthorhombic Pbnm simple perovskite structure, like it is observed in the inset of Fig. 1. The goodness of fit factors for the P2$_1$/n model also validates that structure.

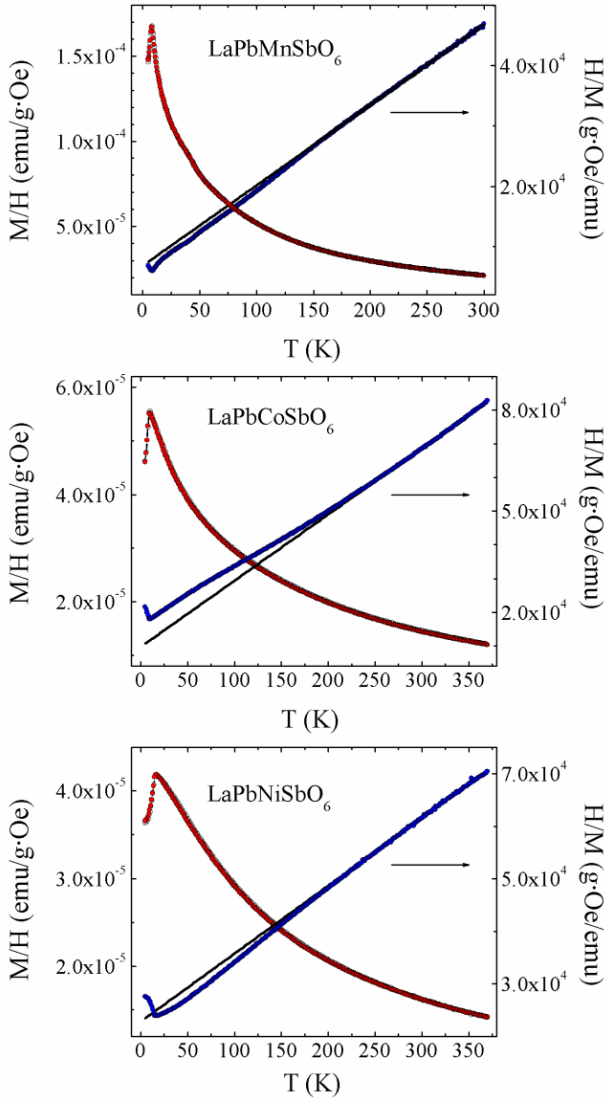

Fig. 2. *M/H* (left axis) and *H/M* (right axis) vs *T* for the double perovskites LaPbMSbO$_6$ measured at 1 kOe, under zero field (white circles) and field cooling (red circles) conditions. The straight lines are the fits obtained with the Curie-Weiss law in the paramagnetic state at high temperature (*T* > 250 K).

**FIG. 2 HERE**

**TABLE I HERE**

| compound | $T_N$ (K) | $T_{CW}$ (K) | $p_e$ (μ$_B$) | $p_t$ (μ$_B$) |
|---|---|---|---|---|
| LaPbMnSbO$_6$ | 8(1) | -56.9(4) | 6.2(1) | 5.92 |
| LaPbCoSbO$_6$ | 10(1) | -62.0(6) | 5.12(2) | 6.63 |
| LaPbNiSbO$_6$ | 17(1) | -119(4) | 5.5(2) | 5.59 |

$T_N$ was defined as the maximum in *M/H* vs *T*. $T_{CW}$ and $p_e$ were extracted from the Curie-Weiss fit of the high temperature zone of the *H/M* vs *T* curves. $p_t$ is the expected magnetic moment for M$^{2+}$, calculated taking into account the orbital contribution and considering high spin electronic configuration for Mn$^{2+}$ and Co$^{2+}$.

The M$^{2+}$ and Sb$^{5+}$ distribution among the *2d* and *2c* octahedral sites was allowed to vary. For M$^{2+}$ = Mn$^{2+}$ and Co$^{2+}$ a maximum order situation was identified, so the formula for these cases is LaPb(M)$_{2d}$(Sb)$_{2c}$O$_6$, whereas for M$^{2+}$ = Ni$^{2+}$ a 3% disorder was found. A high degree of order is expected given the high differences in charge ($\Delta q$) and ionic radii ($\Delta r$) between the transition metal M$^{2+}$ and Sb$^{5+}$. $\Delta q$ is 3 for all the samples, and $\Delta r$ is 0.23, 0.145, and 0.09 Å, for Mn$^{2+}$, Co$^{2+}$ and Ni$^{2+}$ respectively, calculated from the corresponding ionic radii [7]. As can be seen Ni$^{2+}$ has, among the transition metals, the closest ionic radi to Sb$^{5+}$, and these similarity between Ni$^{2+}$ and Sb$^{5+}$ explains the small disorder found for LaPbNiSbO$_6$. The complete set of structural data [4] will be published elsewhere.

Thermal evolution of the magnetization (*M/H* vs *T*) at 1 kOe is showed for the three samples in Fig. 2, where it can be seen that there is no difference between the data obtained under ZFC and FC conditions. At room temperature the double perovskites are paramagnetic, and at low temperatures typical features of an antiferromagnetic order are observed for the three oxides, with Neel temperatures $T_N$ –taken as the maximum in *M/H* vs *T* data- below 20 K, Table I.

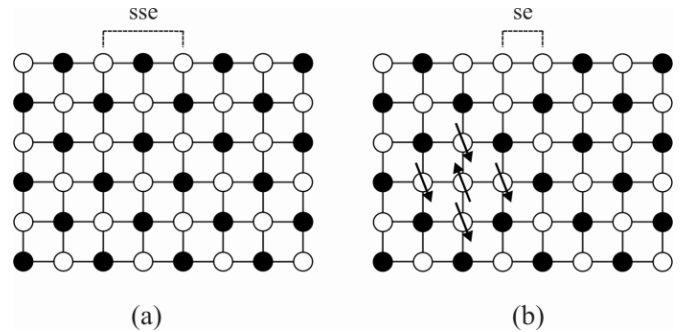

Fig. 3. Schematic two-dimensional representation of the distribution of octahedral ions in a double perovskite. For simplicity a square structure has been supposed, and oxygen and cuboctahedral ions (La$^{3+}$/Pb$^{2+}$) were omitted. The line linking a white circle (M$^{2+}$) with a black circle (Sb$^{5+}$) represents a M$^{2+}$-O-M$^{2+}$ path. (a) Completely ordered situation, with super-superexchange paths only (sse). (b) Structure with a small disorder between M$^{2+}$ and Sb$^{5+}$, which originates some superexchange interactions (se) and ferrimagnetic regions.

**FIG.3 HERE**

Because La$^{3+}$, Pb$^{2+}$ and Sb$^{5+}$ are diamagnetic ions, the magnetic order is originated by paramagnetic M$^{2+}$. Interactions between M$^{2+}$ ions in an ordered double perovskite structure are mediated by super-superexchange paths only, M$^{2+}$-O$^{2-}$-Sb$^{5+}$-O$^{2-}$-M$^{2+}$, as it can be seen in the scheme of Fig. 3(a). This interaction is antiferromagnetic in nature and the long distances between M$^{2+}$ ions, approximately 8.1 Å for these compounds, explain the low Neel temperatures of the samples and other similar systems like SrLaNiSbO$_6$ ($T_N$ = 26 K) [8], LaCaCoNbO$_6$ ($T_N$ = 16 K) [9], and Ba$_2$MnWO$_6$ ($T_N$ = 9 K) [10].

Above 250 K, $H/M$ vs $T$ for LaPbMSbO$_6$ shows a linear behavior, Fig. 2, and was fitted with the Curie-Weiss law: $H/M = T/C + T_{CW}/C$, where $C$ is the Curie constant and $T_{CW}$ the Curie-Weiss temperature. The negative $T_{CW}$ obtained for the three double perovskites, showed in Table I, confirms that the predominant interactions are antiferromagnetic. The Curie constant allows the calculation of the experimental magnetic moments ($p_e$) for each compound, which are also shown in Table I together with the expected values for M$^{2+}$ ions ($p_t$) with orbital contribution and high spin (HS) configuration for Mn$^{2+}$ and Co$^{2+}$ -$t_{2g}^3e_g^2$ and $t_{2g}^5e_g^2$ respectively-. For manganese and nickel the agreement is very good, but for cobalt the experimental value is significantly lower than the expected one. This indicates a partial orbital contribution, as has been previously observed for example in Ba$_3$CoSb$_2$O$_9$ ($p_e$ = 5.23(3) μ$_B$) [11], SrLaCoSbO$_6$ ($p_e$ = 5.14 μ$_B$), SrLaCoNbO$_6$ ($p_e$ = 4.91 μ$_B$) and SrLaCoTaO$_6$ ($p_e$ = 4.76 μ$_B$) [12] and Sr$_2$CoWO$_6$ ($p_e$ = 5.2 μ$_B$) [13].

Besides the antiferromagnetic features just described, a closer inspection of $M/H$ vs $T$ dependence for LaPbMnSbO$_6$, Fig. 2 – top panel, reveals also a small shoulder between 40 and 50 K, not evident in the samples with cobalt and nickel. In Fig. 4(a) we show an enlargement of this region with an additional measure performed under an applied field of 0.5 kOe. At 0.5 kOe the anomaly is more pronounced than at 1 kOe, and also is observed irreversibility between ZFC and FC magnetization, which starts approximately at 60 K. This anomaly is superimposed in this sample to a larger paramagnetic signal. Its shape and decreasing relative magnitude with the increase of the applied field are characteristics features of a ferro/ferrimagnetic arrangement. The Curie temperature of this magnetic order is 45 K, as obtained from the maximum in the magnetization derivative $dM/dT$ vs $T$ (data not shown).

The synthesis, structure and magnetic properties of a similar double perovskite SrLaMnSbO$_6$ [14], which has also a P2$_1$/n structure and cell parameters close to LaPbMnSbO$_6$ ones have been previously reported. The magnetic phenomenology is also similar: SrLaMnSbO$_6$ has a long range antiferromagnetic order with $T_N$ = 8 K, and a rapid increase of the magnetization on the $M/H$ vs $T$ curve below 250 K is observed, although much more pronounced that the one for LaPbMnSbO$_6$ below 50 K. Although the authors claims that $H/M$ vs $T$ data show substantial nonlinearity over the whole temperature range, they fit the 300-400 K region with a Curie-Weiss law. From the extracted positive Curie-Weiss temperature ($T_{CW}$ = 62 K) they conclude there are ferromagnetic interactions, and propose the presence of a small amount of Mn$^{3+}$ which would give rise to Mn$^{2+}$-O-Mn$^{3+}$ superexchange, leading to ferromagnetic clusters.

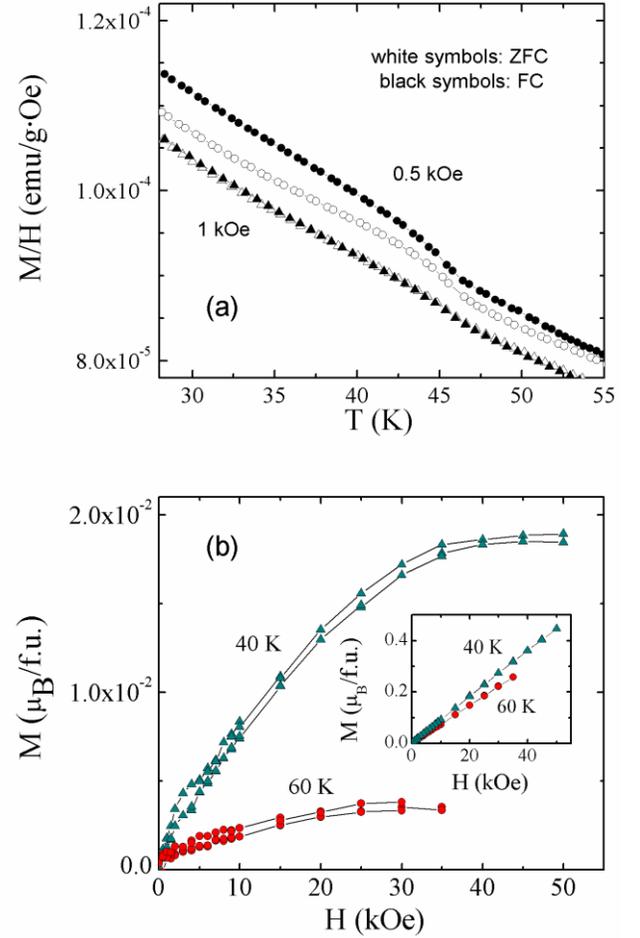

FIG. 4. (a) Magnetization ($M$) vs Temperature ($T$) measured at 0.5 kOe (circles) and 1 kOe (triangles) showing the anomaly at 45 K for LaPbMnSbO$_6$. (b) Ferromagnetic-type dependence of the Magnetization ($M$) vs applied Field ($H$) at 60 and 40 K, obtained subtracting to the total measure (inset) a paramagnetic major component.

**FIG. 4 HERE**

These authors support the presence of Mn$^{3+}$ with two results. First X-ray absorption spectra for Mn – K edge shows the presence of Mn$^{2+}$ formal oxidation state, but also a high energy shoulder which suggest some fraction of Mn$^{3+}$. However, it is said that the shoulder could also be explained by solid-state 4p orbital splitting. On the other hand, neutron powder diffraction was analyzed allowing deviations in chemical composition, resulting in SrLaMn$_{1.05}$Sb$_{0.95}$O$_6$, where a small amount of Mn$^{3+}$ must be necessarily present due to charge balance. Nevertheless nothing is said about oxygen vacancies. Therefore the presence of Mn$^{3+}$ is not clear at all. Probably the positive $T_{CW}$ obtained is not the true Curie-Weiss temperature, and a measurement at higher temperatures than 400 K would allow the access to the real paramagnetic region.

For LaPbMnSbO$_6$ there is no reason to suspect the presence of Mn$^{3+}$, because the paramagnetic region is reached at 250 K (Fig. 2), $T_{CW}$ is negative and no impurity phases were detected in the X-ray diffraction pattern. We think that the microscopic origin of magnetic order at 45 K for LaPbMnSbO$_6$ could be attributed to the distribution of Mn$^{2+}$ and Sb$^{5+}$ over the two octahedral sites. The formula obtained from Rietveld analysis is LaPb(Mn)$_{2d}$(Sb)$_{2s}$O$_6$, which leads to a structure like the one shown in Fig. 3(a), and this is consistent with the antiferromagnetic order at 8 K. However a small disorder could exists, below the detection limit of the refinement. Because of this, proofs with imposed disorders were performed. The results show that 1%, 2%, and even 3% of disorder leads to no significant differences in the quality of the refinements, compared with the case of perfect order. A small disorder would produce a structure like the one in Fig. 3(b). It can be seen that this situation necessarily originates superexchange Mn$^{2+}$-O$^{2-}$-Mn$^{2+}$ paths, also antiferromagnetic according to Goodenough-Kanamori rules for the interaction between two d$^5$ ions (Mn$^{2+}$) [15], but now there is not a total cancelation of the magnetic moments.

This picture gives rise to nanoclusters which are ferrimagnetic, like the Fe-rich patches described for Sr$_3$Fe$_2$UO$_9$ [16], but in this case the regions are far away among them. Upon cooling, and because the characteristic length of superexchange is half the super-superexchange one - that is approximately 4 Å-, the first magnetic interactions that comes into play are Mn$^{2+}$-O$^{2-}$-Mn$^{2+}$ and the few ferrimagnetic regions at 45 K. At lower temperatures Mn$^{2+}$-O$^{2-}$-Sb$^{5+}$-O$^{2-}$-Mn$^{2+}$ interactions starts to become important and finally, at 8 K, lead to an antiferromagnetic order. SrLaMnSbO$_6$ has 2% of disorder [14], and perhaps the ferrimagnetic like phenomenology could also be explained taking into account ferrimagnetic regions.

Looking for additional information about the magnetic order at 45 K, magnetization vs applied field dependences were recorded, inset of Fig. 4(b). At 60 and 40 K the behavior is linear, like a typical paramagnet. However, the subtraction of a line obtained fitting the high field data, leads to a ferrimagnetic like curves, as shown in Fig 4(b). The saturation value is 3.54 x 10$^{-3}$ μ$_B$/f.u. at 60 K and 1.86 x 10$^{-2}$ μ$_B$/f.u. at 40 K. For comparison, the expected saturation for a hypothetic ferromagnetic LaPbMnSbO$_6$ would be 5.92 μ$_B$/f.u., showing that the observed saturations correspond to a small fraction of the sample, consistent with the picture of isolated ferrimagnetic regions.

## IV. CONCLUSION

The double perovskites LaPbMSbO$_6$ (M$^{2+}$ = Mn$^{2+}$, Co$^{2+}$, and Ni$^{2+}$) were obtained by an aqueous synthesis route as pure polycrystals. Order between M$^{2+}$ and Sb$^{5+}$ among the two octahedral sites is higher for Mn$^{2+}$ and Co$^{2+}$ than for Ni$^{2+}$, as obtained from the Rietveld analysis of the X-ray powder diffraction patterns. The magnetization curves show an antiferromagnetic order at temperatures below 20 K for the three samples, as a consequence of the long interaction paths M$^{2+}$-O$^{2-}$-Sb$^{5+}$-O$^{2-}$-M$^{2+}$. However, for LaPbMnSbO$_6$ an additional feature is observed at higher temperatures, 45 K, probably due to a small disorder which originates shorter superexchange Mn$^{2+}$-O$^{2-}$-Mn$^{2+}$ paths and uncompensated isolated regions of ferrimagnetic nature are generated.

## ACKNOWLEDGMENT

R.E.C and G.N are members of CONICET, Argentina. D.G.F. has a scholarship from CONICET, Argentina. Work partially supported by ANPCyT, FONCYT, SECYT-UNC, SeCTyP-UNCuyo and CONICET.